\documentclass[aps,prd,superscriptaddress,10 pt]{revtex4-2}

\usepackage{graphicx}
\usepackage{bm}
\usepackage{booktabs}
\usepackage{xcolor}
\usepackage{orcidlink}
\usepackage{siunitx}
\usepackage{mathtools}
\usepackage{amsmath,amssymb}

\begin{document}

\title{Constraining Ultralight Dark Matter in Dynamical Dark Energy Cosmologies}

\author{Muhammad Awais\,\orcidlink{0009-0000-9174-8492}}
\email{mawais2500@alumno.ipn.mx}
\affiliation{Escuela de F\'isica y Matem\'aticas del Instituto Polit\'ecnico Nacional, Av. Instituto Polit\'ecnico Nacional, San Pedro Zacatenco, 07360, M\'exico}

\author{Hafiz Inam Ullah\,\orcidlink{0009-0005-5796-6054}}
\email{hinamu2500@alumno.ipn.mx}
\affiliation{Escuela de F\'isica y Matem\'aticas del Instituto Polit\'ecnico Nacional, Av. Instituto Polit\'ecnico Nacional, San Pedro Zacatenco, 07360, M\'exico}

\author{Rafael C. Nunes}
\email{rafadcnunes@gmail.com}
\affiliation{Instituto de F\'isica, Universidade Federal do Rio Grande do Sul, 91501-970 Porto Alegre, RS, Brazil}
\affiliation{Divis\~ao de Astrof\'isica, Instituto Nacional de Pesquisas Espaciais, S\~ao Jos\'e dos Campos, SP, Brazil}

\author{Paula S. Ferreira\orcidlink{0000-0002-7540-040X}}
\email{ferreira\_paula@zju.edu.cn}
\affiliation{Institute of Astronomy, School of Physics, Zhejiang University, Hangzhou, China}
\affiliation{Center for Cosmology and Computational Astrophysics, Institute for Advanced Study in Physics, Zhejiang University, Hangzhou, China.}

\author{Tonatiuh Matos}
\email{tonatiuh.matos@cinvestav.mx}
\affiliation{Departamento de F\'isica, Centro de Investigaci\'on y de Estudios Avanzados del Instituto Polit\'ecnico Nacional, Av. Instituto Polit\'ecnico Nacional 2508, San Pedro Zacatenco, 07360 Ciudad de M\'exico, M\'exico}

\date{\today}

\begin{abstract}
We investigate the cosmological implications of scalar-field dark matter (SFDM) in a dynamical dark energy framework, extending previous analyses that considered SFDM within the standard $\Lambda$CDM cosmological model. We consistently implement the scenario at both the background and linear perturbation levels. We perform Bayesian parameter inference using Planck CMB measurements, DESI DR2 baryon acoustic oscillation (BAO) data, and Pantheon+ Type Ia supernovae to constrain the model parameter space.
Our theoretical predictions show that lighter scalar-field masses produce increasingly strong scale-dependent suppression of matter fluctuations and more pronounced modifications to the CMB temperature power spectrum, whereas heavier masses progressively approach the cold dark matter limit. For the SFDM mass, we obtain a one-sided 95\% lower limit of $\log_{10}(m_\phi/{\rm eV})>-24.04$ from our main joint analysis. We find that replacing the standard cold dark matter fluid with SFDM does not qualitatively alter the constraints on dynamical dark energy or the evidence for its evolution, including a phantom crossing at low redshifts.
\end{abstract}

\maketitle

\section{Introduction}

The standard cosmological model provides a successful description of
cosmic microwave background (CMB) anisotropies and the large-scale
distribution of matter \cite{Planck2018}. Evidence for late-time
accelerated expansion was established using Type Ia supernova
distances \cite{Riess1998,Perlmutter1999}. In $\Lambda$CDM, dark
matter is represented by a pressureless cold-dark-matter (CDM) fluid,
while a cosmological constant ($\Lambda$) accounts for the accelerated expansion.
The model does not identify the dark-matter particle, and the
observed value of the cosmological constant raises a longstanding
theoretical question \cite{Weinberg1989}. Moreover, the value of $H_0$ inferred from CMB and BAO observations within the $\Lambda$CDM framework differs from that obtained from local, model-independent measurements (see \cite{CosmoVerse2025} for a review). Assessing this discrepancy requires accounting for the cosmological assumptions underlying each determination and the observational data used in the inference.

The physical nature of dark energy (DE), which drives the accelerated expansion of the Universe, remains an open question in modern cosmology. Within the standard $\Lambda$CDM framework, cosmic acceleration is attributed to the cosmological constant, characterized by an equation-of-state (EoS) parameter $w=-1$ and a constant energy density. However, recent results from the Dark Energy Spectroscopic Instrument (DESI) have renewed interest in departures from this standard picture. Using baryon acoustic oscillation (BAO) measurements from an unprecedented sample of more than 14 million galaxies and quasars, the second DESI data release has reported indications of deviations from $\Lambda$CDM, with statistical significance reaching up to $4\sigma$, depending on the combination of datasets considered \cite{DESIDR2,DESIDR2Extended}. These findings motivate further investigation into dynamical dark energy models, in which the DE density and equation of state may evolve over cosmic time, challenging the conventional interpretation of DE as a pure cosmological constant. Analyses using galaxy clustering, galaxy--CMB lensing cross-correlations, cosmography, and combinations of distance and growth measurements have reported model- and dataset-dependent preferences for evolving dark energy \cite{SilvaNunes2025,Rippentrop2026,deCruzPerez2025,Efstratiou2025,XuChen2025,Fazzari2025,IshakMedinaVarela2025,SabogalNunes2025,LuSimon2025,SmithEtAl2025}. Other analyses find weaker or no statistically significant preference \cite{XuKumar2026,Chudaykin2026,Chaudhary2025}, while alternative physical interpretations and methods for constructing dark-energy models have also been explored \cite{Liu2025Axion,Guedezounme2026,Bom2026}.

On the other hand, scalar-field dark matter (SFDM) provides an alternative description of the dark matter component, in which an ultralight, spinless particle is represented by a scalar field. Early studies of scalar-field cosmologies explored the role of such fields in the cosmic energy budget and structure formation \cite{MatosGuzmanUrena1999,MatosUrena2001,Matos:1998vk,Hu2000,UrenaGonzalez2016,UrenaLopez2024,ZhangEtAl2024Birefringence,LiuHuGrin2025,ZelmerEtAl2025,GhoshBoddyYu2025,LagueEtAl2026,Armengaud:2017nkf,Hlozek:2014lca,Magana:2012xe,Verdiani:2026cpc,Liu:2026jkq,Aboubrahim:2024spa,Moss:2025ymr,Tsutsui:2022zos}. \footnote{An extension of this investigation to the case of spin-1 ultralight dark matter has recently been explored in the literature \cite{FerreiraChaseLopezNacir2024,FerreiraChaseEtAl2025,FerreiraChaseLopezNacir2026,AhumadaAcunaEtAl2026}}. Since then, this framework has been revisited under different names, including fuzzy dark matter, wave dark matter, and ultralight dark matter (see \cite{Ferreira2021ULDM,EberhardtFerreira2025} for a review). In particular, the wave nature of the field can suppress structure formation below a characteristic scale, with potentially significant implications for cosmological and astrophysical observables \cite{Hu2000,Schive2014,Marsh2016,Hui2017}. Subsequent studies have investigated the phenomenological implications of SFDM across a wide range of astronomical probes, highlighting its potential to address several questions in cosmology and astrophysics \cite{Muller:2018hks,Bernal:2024wif,Hernandez-Marquez:2026lkw,Matos:2022ruq,Matos:2023usa,Farren:2021jcd,Gaughan:2026xrv,Lopez-Sanchez:2025uwk,Liu:2025qwf,Ghosh:2025pbn,Furlanetto:2024qsb,Verdiani:2026cpc}. However, these investigations consider different scalar-field potentials and modeling assumptions, and their results must therefore be interpreted within the context of the specific SFDM scenario under consideration.

In this work, we considered SFDM  as a minimally coupled,
canonical real field with the quadratic potential
$V(\phi)=m_\phi^2\phi^2/2$. When $m_\phi$ is large compared with the
Hubble expansion rate, rapid oscillations make the field's
time-averaged background pressure approach zero. Its perturbations
retain a mass-dependent gradient response, however, which changes
the growth of structure below a characteristic Jeans scale.  Simulations have investigated
further consequences of wave dynamics inside halos
\cite{Schive2014}, while Lyman-$\alpha$ forest measurements have been
used to constrain small-scale suppression subject to assumptions
about the intergalactic medium \cite{Irsic2017}. A mass constraint from any such analysis must be associated with its field model-dependent, observables, and likelihoods.

For the late-time dark energy sector, we consider the VCDM model, in which the cosmological evolution is governed by a nondynamical auxiliary field \cite{DeFelice2020}. Previous studies have investigated modifications to the low-redshift expansion history within this framework and their observational consequences \cite{DeFelice2021} (see also ~\cite{Arora2025VCDM, Borghetto2026VCDM}). More recently, Ref.~\cite{Scherer2025} introduced a smooth late-time transition in two settings: a phenomenological $w_\dagger$CDM model with CDM and its extension to $w_\dagger$VCDM. This transition is characterized by its amplitude, epoch, and width, allowing the DE equation-of-state parameter to cross the phantom divide, $w=-1$. This framework has demonstrated the potential to accommodate recent observational results and provides a physically motivated alternative to commonly used phenomenological parametrizations of dynamical dark energy. 
Without loss of generality, we do not consider additional parametrizations, as the adopted model provides sufficient flexibility to achieve an excellent fit to the observational datasets considered in this work, while allowing us to investigate the effects of SFDM in the presence of dynamical dark energy.

These two transition models serve as reference cases for the present work. Replacing CDM in $w_\dagger$VCDM with SFDM raises a fundamental question: do the mass-dependent dynamics of SFDM alter the inferred properties of the DE transition and the predicted clustering observables? Since the VCDM scalar constraint depends on the matter enthalpy and momentum, addressing this question requires a consistent treatment of SFDM beyond simply replacing the CDM background density \cite{DeFelice2020,DeFelice2021,Scherer2025}.

Our objective is to construct and investigate a cosmological framework that combines SFDM with dynamical DE, described by a transition model capable of providing an excellent fit to current observational data. We examine how the scalar-field mass $m_\phi$ and the properties of the late-time dark energy transition affect the matter power spectrum and CMB temperature anisotropies. We then constrain the full model using \textit{Planck} 2018 temperature and polarization measurements, \textit{BAO DESI DR2} data, and Type Ia supernova observations from the Pantheon+ sample. Finally, we discuss the resulting parameter constraints and their implications for the interplay between the nature of DM and the late-time evolution of DE.

\section{Cosmological Framework}

In this section, we review the theoretical aspects of the cosmological model proposed in this work.

\subsubsection{Background evolution}

Following Ref. \cite{UrenaLopez2024}, we model SFDM as a minimally coupled, canonical real scalar field
$\phi$. In a spatially flat Friedmann--Lema\^itre--Robertson--Walker
background, the line element is
\begin{equation}
 ds^2=-dt^2+a^2(t)\delta_{ij}\,dx^i dx^j,
 \label{eq:sfdm_metric}
\end{equation}
where $H\equiv\dot a/a$ and overdots denote derivatives with respect to cosmic time.
For the combined model, we write the effective background density as
\begin{equation}
 \rho_{\rm tot}=\rho_\gamma+\rho_\nu+\rho_b+\rho_\phi+\rho_x,
 \label{eq:sfdm_total_density}
\end{equation}
where $\rho_x$ denotes the effective late-time VCDM component
specified in the following subsection. Here $\rho_\nu$ includes both relativistic and massive neutrinos, and
$p_{\rm tot}$ includes the effective pressure $p_x=w_x\rho_x$.
With $\kappa^2\equiv8\pi G$, the background equations take the form
\begin{align}
 H^2&=\frac{\kappa^2}{3}\rho_{\rm tot},
 \label{eq:sfdm_friedmann}\\
 \dot H&=-\frac{\kappa^2}{2}
 (\rho_{\rm tot}+p_{\rm tot}).
 \label{eq:sfdm_hdot}
\end{align}

The scalar-field action is
\begin{equation}
 S_\phi=\int d^4x\,\sqrt{-g}
 \left[
 -\frac{1}{2}g^{\mu\nu}\partial_\mu\phi\partial_\nu\phi
 -V(\phi)
 \right].
 \label{eq:sfdm_action}
\end{equation}
For this work, we choose the quadratic potential
\begin{equation}
 V(\phi)=\frac{1}{2}m_\phi^2\phi^2.
 \label{eq:sfdm_potential}
\end{equation}
We use units with $c=\hbar=1$. 

Variation of the action in the homogeneous background gives
\begin{equation}
 \ddot\phi+3H\dot\phi+m_\phi^2\phi=0.
 \label{eq:sfdm_kg}
\end{equation}
The scalar-field energy density and pressure are
\begin{subequations}\label{eq:sfdm_density_pressure}
\begin{align}
 \rho_\phi&=\frac{1}{2}\dot\phi^2
             +\frac{1}{2}m_\phi^2\phi^2,
 \label{eq:sfdm_density}\\
 p_\phi&=\frac{1}{2}\dot\phi^2
          -\frac{1}{2}m_\phi^2\phi^2.
 \label{eq:sfdm_pressure}
\end{align}
\end{subequations}
Consequently, $\rho_\phi+p_\phi=\dot\phi^2$, and its instantaneous
equation-of-state parameter is
\begin{equation}
 w_\phi=\frac{p_\phi}{\rho_\phi}.
 \label{eq:sfdm_w_definition}
\end{equation}
For potential-dominated initial conditions with $H\gg m_\phi$,
Hubble friction leaves the field nearly frozen, so $w_\phi\simeq-1$.
Once $m_\phi\gg H$, rapid oscillations give
\begin{equation}
 \langle w_\phi\rangle\simeq0,
 \qquad
 \langle\rho_\phi\rangle\propto a^{-3}.
 \label{eq:sfdm_averaged_limit}
\end{equation}
This averaged background behavior does not eliminate the
scale dependence of the perturbations \cite{Hu2000,UrenaLopez2024}.

\subsubsection{Dynamical-system formulation}

Defining the reduced Planck mass by $M_{\rm Pl}^{-2}=\kappa^2$, we
introduce
\begin{equation}
 x\equiv\frac{\dot\phi}{\sqrt{6}HM_{\rm Pl}},
 \qquad
 y\equiv-\frac{m_\phi\phi}{\sqrt{6}HM_{\rm Pl}}.
 \label{eq:sfdm_xy}
\end{equation}
The minus sign in $y$ is required by the polar-variable convention
used below. The density fraction is
\begin{equation}
 \Omega_\phi=x^2+y^2.
 \label{eq:sfdm_xy_omega}
\end{equation}
With $N\equiv\ln a$ and a prime denoting $d/dN$, the
Klein--Gordon equation becomes
\begin{align}
 x'&=-3x+\frac{m_\phi}{H}y-x\frac{H'}{H},
 \label{eq:sfdm_xprime}\\
 y'&=-\frac{m_\phi}{H}x-y\frac{H'}{H}.
 \label{eq:sfdm_yprime}
\end{align}
For numerical evolution through the rapidly oscillating regime,
we use the variables of Ref.~\cite{UrenaLopez2024}:
\begin{align}
 \frac{\kappa\dot\phi}{\sqrt{6}H}
   &=e^\beta\sin\frac{\theta}{2},
 \nonumber\\
 -\frac{\kappa m_\phi\phi}{\sqrt{6}H}
   &=e^\beta\cos\frac{\theta}{2},
 \qquad y_1=\frac{2m_\phi}{H}.
 \label{eq:sfdm_variables}
\end{align}
Thus $x=e^\beta\sin(\theta/2)$ and
$y=e^\beta\cos(\theta/2)$. In particular,
\begin{equation}
 \Omega_\phi\equiv\frac{\kappa^2\rho_\phi}{3H^2}
 =e^{2\beta},
 \qquad
 w_\phi=-\cos\theta.
 \label{eq:sfdm_omega_w}
\end{equation}
The equivalent background equations are
\begin{align}
 \frac{d\theta}{dN}&=-3\sin\theta+y_1,
 \nonumber\\
 \frac{dy_1}{dN}&=\frac{3}{2}(1+w_{\rm tot})y_1,
 \nonumber\\
 \frac{d\beta}{dN}&=\frac{3}{2}(w_{\rm tot}+\cos\theta),
 \label{eq:sfdm_background}
\end{align}
where $w_{\rm tot}\equiv p_{\rm tot}/\rho_{\rm tot}$ includes the
effective background components. The variable $\theta$ follows
the oscillatory phase, while $\beta$ determines the instantaneous
SFDM density fraction.

\subsubsection{Linear perturbations}

In synchronous gauge, the perturbed spacetime metric is given by
\begin{equation}
 ds^2=-dt^2+a^2(t)(\delta_{ij}+h_{ij})\,dx^i dx^j,
 \label{eq:sfdm_sync_metric}
\end{equation}
where $a(t)$ is the scale factor and $h_{ij}$ describes the spatial metric perturbations. We decompose the scalar field into its homogeneous background and first-order perturbation,
\begin{equation}
 \phi(\boldsymbol{x},t)=\phi(t)+\delta\phi(\boldsymbol{x},t).
 \label{eq:sfdm_field_decomposition}
\end{equation}
Working in Fourier space, a mode with comoving wavenumber $k$ obeys the linearized Klein--Gordon equation \cite{MaBertschinger1995,UrenaLopez2024},
\begin{equation}
 \delta\ddot\phi+3H\delta\dot\phi+
 \left(m_\phi^2+\frac{k^2}{a^2}\right)\delta\phi
 +\frac{1}{2}\dot\phi\,\dot h=0,
 \label{eq:sfdm_perturbed_kg}
\end{equation}
where $m_\phi$ is the scalar-field mass, and $h$ denotes the trace of the scalar metric perturbation in the standard synchronous-gauge convention. 

The density and pressure perturbations
are
\begin{align}
 \delta\rho_\phi
 &=\dot\phi\,\delta\dot\phi+m_\phi^2\phi\,\delta\phi,
 \label{eq:sfdm_delta_rho}\\
 \delta p_\phi
 &=\dot\phi\,\delta\dot\phi-m_\phi^2\phi\,\delta\phi.
 \label{eq:sfdm_delta_p}
\end{align}
In the equivalent first-order formulation of
Ref.~\cite{UrenaLopez2024}, $\delta_0$ is the SFDM density contrast,
$\delta_\phi=\delta_0=\delta\rho_\phi/\rho_\phi$. Its accompanying
dynamical perturbation variable is
\begin{equation}
 \delta_1=\frac{m_\phi}{\rho_\phi}
 \left(\dot\phi\,\delta\phi-\phi\,\delta\dot\phi\right).
 \label{eq:sfdm_delta1_definition}
\end{equation}
The scalar momentum contribution is
\begin{equation}
 (\rho_\phi+p_\phi)\Theta_\phi
 =\frac{k^2\rho_\phi}{aHy_1}
 \left[(1-\cos\theta)\delta_1-\sin\theta\,\delta_0\right].
 \label{eq:sfdm_momentum}
\end{equation}
The perturbation system used with these variables is
\begin{align}
 \frac{d\delta_0}{dN}
 ={}&-3\sin\theta\,\delta_1
 -\frac{k^2}{k_J^2}(1-\cos\theta)\delta_1
 \nonumber\\
 &+\frac{k^2}{k_J^2}\sin\theta\,\delta_0
 -\frac{1}{2}\frac{dh}{dN}(1-\cos\theta),
 \nonumber\\
 \frac{d\delta_1}{dN}
 ={}&-3\cos\theta\,\delta_1
 -\frac{k^2}{k_J^2}\sin\theta\,\delta_1
 \nonumber\\
 &+\frac{k^2}{k_J^2}(1+\cos\theta)\delta_0
 -\frac{1}{2}\frac{dh}{dN}\sin\theta,
 \label{eq:sfdm_perturbations}
\end{align}
where the characteristic Jeans wavenumber is
\begin{equation}
 k_J^2=a^2H^2y_1=2a^2m_\phi H.
 \label{eq:sfdm_kj}
\end{equation}
The factors $k^2/k_J^2$ give the mass-dependent,
scale-dependent linear response. In the nonrelativistic and
rapidly oscillating limit, an approximate effective-fluid
description has
\begin{equation}
 c_{\rm eff}^2(k,a)\simeq\frac{k^2}{4a^2m_\phi^2}.
 \label{eq:sfdm_sound_speed}
\end{equation}
Equation~(\ref{eq:sfdm_sound_speed}) describes a limiting
effective-fluid behavior and is not used to close the perturbation
system. Numerical evolution instead follows the polar perturbation
variables, with the rapid-oscillation cutoff prescription of
Ref.~\cite{UrenaLopez2024}. In the combined model,
Eqs.~(\ref{eq:sfdm_density_pressure}) and
(\ref{eq:sfdm_momentum}) provide the SFDM contributions to the
background density, enthalpy, and scalar momentum.

\subsection{Dynamical dark energy model}

We follow Ref.~\cite{Scherer2025} and specify the late-time
background by
\begin{equation}
 w_x(N)=-1+\Delta\tanh\!\left[\zeta(N_\dagger-N)\right],
 \qquad N_\dagger=\ln a_\dagger,
 \label{eq:wdagger}
\end{equation}
where $a_\dagger=(1+z_\dagger)^{-1}$ is the transition scale factor,
$\Delta$ its amplitude, and $\zeta$ controls its width. With this
convention, $\Delta<0$ gives $w_x<-1$ before the transition and 
$w_x>-1$ afterward. The effective background density satisfies
\begin{equation}
 \frac{d\rho_x}{dN}=-3(1+w_x)\rho_x.
 \label{eq:x_continuity}
\end{equation}
Normalizing its integral at $a=1$ yields
\begin{equation}
 \rho_x(a)=\rho_{x0}
 \left[
 \frac{\cosh\!\left(\zeta\ln[a/a_\dagger]\right)}
      {\cosh\!\left(\zeta\ln[1/a_\dagger]\right)}
 \right]^{3\Delta/\zeta}.
 \label{eq:x_density}
\end{equation}
For $\Delta=0$, this expression gives $w_x=-1$ and
$\rho_x=\rho_{x0}$, independently of $a_\dagger$. The component
$x$ specifies an effective expansion history; in VCDM it does not
represent an additional propagating dark-energy fluid
\cite{Scherer2025}.

At linear order the VCDM scalar momentum constraint differs from
its general-relativistic counterpart, although the standard matter
degrees of freedom remain present \cite{Scherer2025}. To state the
constraint without confusing physical and rescaled pressures, define
\begin{equation}
 \varrho_I\equiv\frac{\rho_I}{3M_{\rm Pl}^2},
 \qquad
 \bar p_I\equiv\frac{p_I}{3M_{\rm Pl}^2},
 \qquad M_{\rm Pl}^{-2}=8\pi G.
 \label{eq:vcdm_rescaling}
\end{equation}
In the remainder of this subsection a prime denotes a derivative
with respect to conformal time $\tau$, whereas the prime in
Eqs.~(\ref{eq:sfdm_xprime}) and (\ref{eq:sfdm_yprime}) denotes
$d/dN$. We write $\mathcal H=a'/a=aH$ and
$H=a'/a^2$, retaining $H$ for the cosmic-time Hubble rate. Define
\begin{equation}
 R=\sum_{I\ne x}(\varrho_I+\bar p_I),
 \qquad
 Q_{\rm N}=\sum_{I\ne x}
       (\varrho_I+\bar p_I)\theta_{I,{\rm N}}.
 \label{eq:vcdm_sums}
\end{equation}
The sums include baryons, radiation, neutrinos, and dark matter,
but exclude the effective $x$ component. The label N on a velocity
divergence denotes conformal Newtonian gauge.

The replacement of CDM by SFDM must be made in both sums. In the
synchronous gauge used for the field evolution, its contributions are
\begin{align}
 R_\phi&=\varrho_\phi+\bar p_\phi
          =\frac{\dot\phi^2}{3M_{\rm Pl}^2},
 \nonumber\\
 Q_{\phi,{\rm S}}
 &=R_\phi\theta_{\phi,{\rm S}}
  =\frac{k^2\dot\phi\,\delta\phi}{3M_{\rm Pl}^2a}.
 \label{eq:sfdm_vcdm_sources}
\end{align}
Here $\theta_{\phi,{\rm S}}$ denotes the velocity divergence
$\Theta_\phi$ in Eq.~(\ref{eq:sfdm_momentum}). Thus
$R=R_{\rm other}+R_\phi$ and
$Q_{\rm S}=Q_{{\rm other},{\rm S}}+Q_{\phi,{\rm S}}$;
the CDM density parameter is set to zero in the input configuration
of the combined model \cite{UrenaLopez2024,Scherer2025}.

The polar-variable treatment applies a cutoff to rapidly oscillating
functions after the onset of coherent oscillations
\cite{UrenaLopez2024}. In this regime the corresponding averaged
sources are $R_\phi\simeq\varrho_\phi$ and
\begin{equation}
 Q_{\phi,{\rm S}}\simeq
 \frac{k^2\varrho_\phi}{aHy_1}\,\delta_1.
 \label{eq:sfdm_vcdm_averaged_sources}
\end{equation}
The VCDM sums use the same SFDM enthalpy and momentum prescription
as the rest of the perturbation solver; dividing the momentum density
by the instantaneous $\rho_\phi+p_\phi$ would be ill conditioned
when the latter vanishes during an oscillation
\cite{UrenaLopez2024}.

In conformal Newtonian gauge, the VCDM momentum constraint reads
\cite{Scherer2025}
\begin{equation}
 \Phi'+\mathcal H\Psi=
 \frac{3\left(k^2-3aH'\right)Q_{\rm N}}
 {k^2\left(2k^2/a^2+9R\right)},
 \label{eq:vcdm_constraint}
\end{equation}
where $\Phi$ and $\Psi$ are the scalar metric potentials. Since
the SFDM equations above use synchronous gauge, we introduce
\begin{equation}
 \alpha\equiv\frac{h'+6\eta'}{2k^2}.
 \label{eq:alpha}
\end{equation}
The linear gauge transformation \cite{MaBertschinger1995} gives
\begin{align}
 \Phi&=\eta-\mathcal H\alpha,
 &
 \Psi&=\alpha'+\mathcal H\alpha,
 \nonumber\\
 \theta_{I,{\rm N}}&=\theta_{I,{\rm S}}+k^2\alpha.
 \label{eq:gauge_relations}
\end{align}
It follows that $Q_{\rm N}=Q_{\rm S}+k^2R\alpha$ and
\begin{equation}
 \Phi'+\mathcal H\Psi=\eta'-aH'\alpha.
 \label{eq:gauge_constraint_identity}
\end{equation}
These relations put Eq.~(\ref{eq:vcdm_constraint}) in the same
gauge as the SFDM perturbation system.

For clarity, let $\eta'_{\rm GR}=3a^2Q_{\rm S}/(2k^2)$ denote
the standard synchronous momentum source evaluated with the same
background and matter perturbations, and let $h'_{\rm GR}$ be its
corresponding reference metric derivative. Set
$\alpha_{\rm GR}=(h'_{\rm GR}+6\eta'_{\rm GR})/(2k^2)$ and define
\begin{align}
 B&=\frac{2k^2}{a^2}+9R,
 \nonumber\\
 E&=aH'+\frac{3}{2}a^2R
   =-\frac{3}{2}a^2(\varrho_x+\bar p_x).
 \label{eq:vcdm_EB}
\end{align}
The final equality uses the effective background evolution, so
$E=-3a^2(1+w_x)\varrho_x/2$. Substitution of
Eqs.~(\ref{eq:gauge_relations}) and
(\ref{eq:gauge_constraint_identity}) into
Eq.~(\ref{eq:vcdm_constraint}) gives the synchronous-gauge update
\begin{align}
 \eta'&=\eta'_{\rm GR}+\delta\eta',
 \nonumber\\
 h'&=h'_{\rm GR}-6\delta\eta',
 \nonumber\\
 \delta\eta'
 &=\frac{E}{B}
 \left[
   \frac{2k^2}{a^2}\alpha_{\rm GR}
   -\frac{9Q_{\rm S}}{k^2}
 \right].
 \label{eq:metric_update}
\end{align}
The paired corrections leave $h'+6\eta'$ unchanged; hence
$\alpha=\alpha_{\rm GR}$ while the modified momentum constraint
is enforced. They also identify a continuous limiting case:
$\Delta=0$ implies $\varrho_x+\bar p_x=0$, so that $E=0$ and
$\delta\eta'=0$.

\subsection{Combined SFDM and VCDM implementation}

A consistent CMB analysis requires evolving the scalar field at both the background and linear-perturbation levels. Dynamical-variable methods enable the treatment of rapidly oscillating quadratic fields without explicitly resolving their increasingly rapid oscillations \cite{UrenaGonzalez2016}. The accuracy of this approach was assessed in a modified \texttt{CLASS} implementation by \citet{UrenaLopez2024}, considering the background evolution, linear perturbations, CMB anisotropies, and matter power spectrum. Similarly, \citet{TellezTovar2022} investigated single- and multiple-field SFDM scenarios, including the quadratic potential, using a modified version of \texttt{CLASS}. We adopt their single-field quadratic formulation, as further developed in Ref.~\cite{UrenaLopez2024}, and extend it to the late-time VCDM framework. In this unified description, the scalar field replaces the CDM fluid in both the background evolution and the linear-perturbation equations.

Our implementation is based on a modified version of \texttt{CLASS}\footnote{Public \texttt{CLASS} repository: \url{https://github.com/lesgourg/class_public}.} \cite{CLASS}, combining the SFDM formulation of Refs.~\cite{TellezTovar2022,UrenaLopez2024} with the late-time transition and VCDM constraint prescription of Ref.~\cite{Scherer2025}. This framework allows us to compare three models: $w_\dagger$CDM, which retains CDM and incorporates a phenomenological transition; $w_\dagger$VCDM, which retains CDM while implementing the VCDM scalar constraint; and SFDM+$w_\dagger$VCDM, in which the quadratic scalar field replaces CDM within the latter framework.

The replacement is implemented at the background and perturbation
levels. The background contains $\rho_\phi$ and $p_\phi$ alongside
the effective $\rho_x(a)$ of Eq.~(\ref{eq:x_density}). For each
linear Fourier mode, the SFDM variables provide
$\delta\rho_\phi$ and $Q_{\phi,{\rm S}}$; the latter enters the
complete $R$ and $Q_{\rm S}$ sums of
Eqs.~(\ref{eq:vcdm_sums}) and
(\ref{eq:sfdm_vcdm_sources}), using the averaged source
(\ref{eq:sfdm_vcdm_averaged_sources}) when appropriate. The resulting
correction (\ref{eq:metric_update}) changes the synchronous metric
sources that enter the SFDM and other species' perturbation equations.
The effective $x$ component is specified by the background
transition and the VCDM constraint, rather than by an independently
propagating dark-energy-fluid perturbation. In particular, the
VCDM auxiliary sector and the physical matter field $\phi$ play
different roles: the latter carries a density perturbation and
momentum density, while the former changes the scalar gravitational
constraint \cite{Scherer2025}.

Two limits clarify the relation to the comparison models. Setting
$\Delta=0$ makes $\rho_x$ constant and removes the VCDM correction
in Eq.~(\ref{eq:metric_update}), leaving quadratic SFDM with a
cosmological-constant background. When the field oscillates rapidly
and $k^2/k_J^2\ll1$ on the scales of interest, its averaged
background and large-scale perturbations approach those of CDM;
the combined model then approaches the corresponding
CDM-based transition history on those scales
\cite{Hu2000,UrenaLopez2024}.

\section{Datasets and methodology}
\label{sec:data_methods}

To investigate the cosmological predictions of SFDM in the dynamical DE framework, we perform a Bayesian parameter inference analysis using cosmic microwave background (CMB), baryon acoustic oscillation (BAO), and Type Ia supernova (SNIa) observations. We consider two combinations of observational datasets: \textit{Planck+BAO DESI DR2}, comprising Planck 2018 CMB temperature and polarization measurements combined with the Dark Energy Spectroscopic Instrument (DESI) Data Release 2 BAO measurements, and \textit{Planck+BAO DESI DR2+PP}, which additionally includes the Pantheon+ SNIa sample. We constrain the parameters of the SFDM+$w_\dagger$VCDM model and compare the results with those obtained for the $w_\dagger$CDM and $w_\dagger$VCDM scenarios.

For the CMB, we use the Planck 2018 temperature and polarization likelihoods. At high multipoles, we adopt the Plik TT, TE, and EE likelihoods, covering $30\leq\ell\leq2508$ for TT and $30\leq\ell\leq1996$ for TE and EE. At low multipoles, we include the TT and EE likelihoods over $2\leq\ell\leq29$. These measurements provide constraints on the primordial perturbations and the background cosmological parameters \cite{Planck2018Overview,Planck2018Likelihoods}.

For BAO, we use the \textit{BAO DESI DR2} measurements from galaxies, quasars, and the Lyman-$\alpha$ forest. These observations constrain cosmological distances relative to the sound horizon at the baryon drag epoch. We account for the correlations between the BAO measurements using the covariance matrices provided with the corresponding likelihoods \cite{DESIDR2,DESIDR2LyA}.

Finally, for the \textit{Planck+BAO DESI DR2+PP} combination, we incorporate the Pantheon+ SNIa dataset, which comprises 1701 light curves from 1550 distinct supernovae \cite{PantheonPlus}. These observations constrain the late-time expansion history through luminosity-distance measurements. We use the Pantheon+ likelihood without the SH0ES distance-anchor calibration, thereby avoiding the inclusion of an external local calibration of the Hubble constant.

The background evolution and linear cosmological perturbations of
SFDM+$w_\dagger$VCDM are computed using our modified version of the Boltzmann
solver \texttt{CLASS}. 
The implementation combines the single-field quadratic SFDM sector with the $w_\dagger$VCDM
transition, with SFDM replacing the cold-dark-matter fluid. The scalar-field
energy density, pressure, and linear perturbations are consistently included
in the background and perturbation evolution
\cite{UrenaLopez2024,TellezTovar2022}.
We perform Bayesian parameter inference using the Markov chain Monte Carlo
(MCMC) sampler implemented in \texttt{Cobaya}\footnote{The public
\texttt{Cobaya} repository is available at
\url{https://github.com/CobayaSampler/cobaya}.} \cite{Cobaya}.
The marginalized one- and two-dimensional posterior distributions are
analyzed with \texttt{GetDist}\footnote{The public \texttt{GetDist} repository
is available at \url{https://github.com/cmbant/getdist}.} \cite{GetDist2025}.

The SFDM+$w_\dagger$VCDM cosmological parameter space contains the nine
parameters $\{10^2\omega_b,\Omega_{\phi,0},100\theta_s,
\ln(10^{10}A_s),n_s,\tau_{\rm reio},\Delta,z_\dagger,
\log_{10}(m_\phi/{\rm eV})\}$. Here
$\omega_b\equiv\Omega_bh^2$ denotes the physical baryon density,
$\Omega_{\phi,0}$ is the present-day SFDM density fraction,
$\theta_s$ is the angular size of the sound horizon at last scattering,
$A_s$ and $n_s$ specify the amplitude and scalar spectral index of the
primordial power spectrum, and $\tau_{\rm reio}$ is the optical depth to
reionization. The parameters $\Delta$ and $z_\dagger$ describe the
late-time transition, with $z_\dagger=a_\dagger^{-1}-1$.
We adopt flat priors
$10^2\omega_b\in[0.5,10.0]$,
$\Omega_{\phi,0}\in[0,1]$,
$100\theta_s\in[0.5,2.0]$,
$\ln(10^{10}A_s)\in[1,5]$,
$n_s\in[0.1,2]$,
$\tau_{\rm reio}\in[0.004,0.8]$,
$\Delta\in[-5,5]$,
$z_\dagger\in[0,10]$, and
$\log_{10}(m_\phi/{\rm eV})\in[-27,-17]$.

For the SFDM+$w_\dagger$VCDM model, we set $\omega_{\rm cdm}=0$ and assume a spatially flat universe, $\Omega_k=0$. The radiation sector follows the standard setup, with two massless neutrino species and one massive neutrino species with a mass of $0.06\,\mathrm{eV}$. We adopt $T_{\rm CMB}=2.7255\,\mathrm{K}$, impose BBN consistency for the primordial helium abundance, use \texttt{RECFAST} for recombination, and assume adiabatic scalar initial conditions. The linear perturbations are evolved in synchronous gauge. We also derive the clustering parameter $S_8\equiv\sigma_8\sqrt{\Omega_m/0.3}$.

We monitor MCMC convergence using the Gelman--Rubin diagnostic, with the
configured convergence threshold $R - 1 < 0.02$ \cite{GelmanRubin1992}.
The marginalized constraints and two-dimensional credible regions are
obtained with \texttt{GetDist} \cite{GetDist2025}. 

\begin{figure*}[tbp]
  \centering
\includegraphics[width=0.91\textwidth,height=0.76\textheight,keepaspectratio]{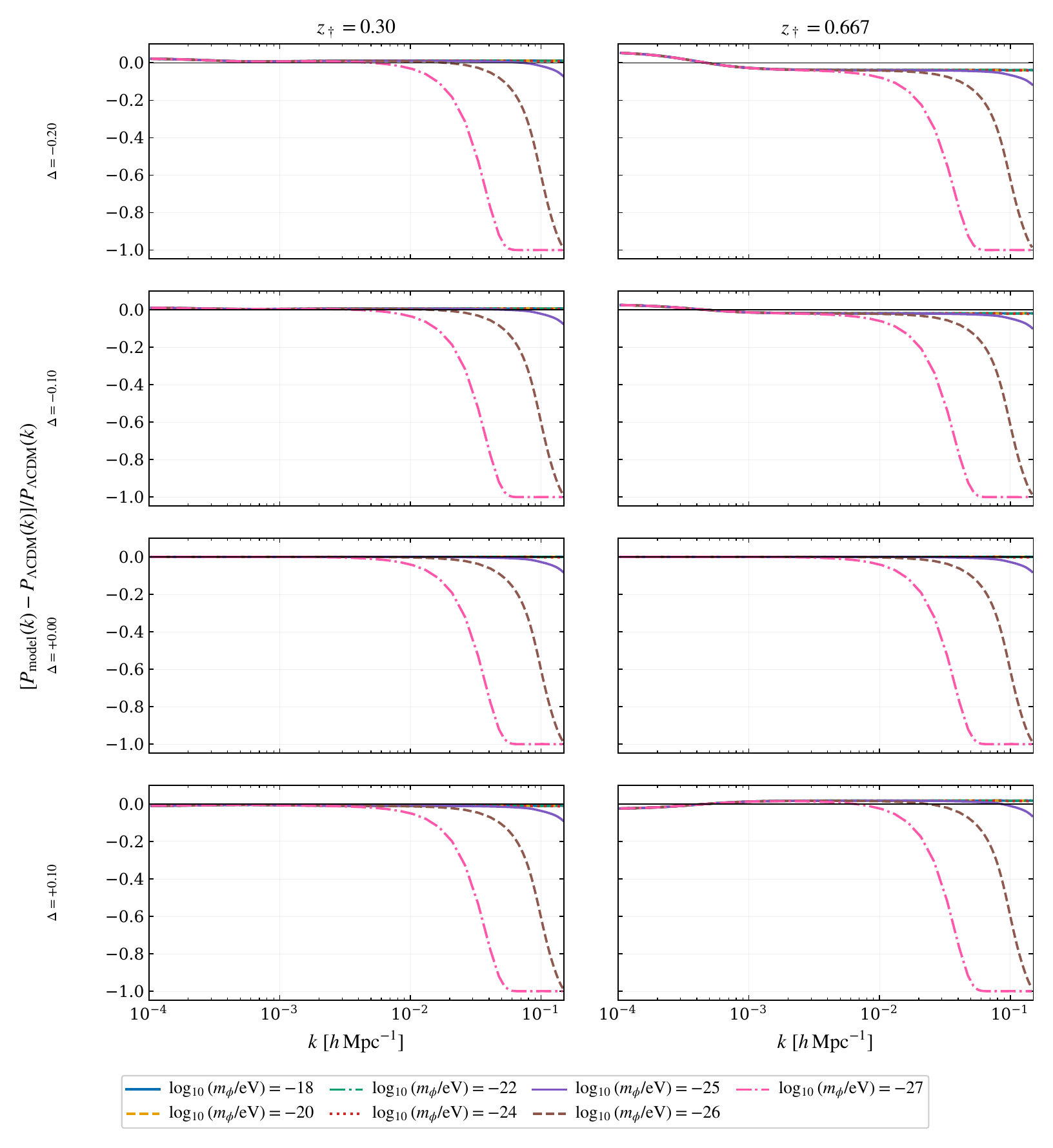}
  \caption{Fractional difference between the linear matter power spectra
  of SFDM+$w_\dagger$VCDM and the $\Lambda$CDM reference model,
  $[P_{\rm model}(k)-P_{\Lambda\mathrm{CDM}}(k)]/
  P_{\Lambda\mathrm{CDM}}(k)$ as a function of comoving wavenumber
  $k$ in $h\,{\rm Mpc}^{-1}$. Columns correspond to
  $z_\dagger=0.30$ (left) and $z_\dagger=0.667$ (right); rows show
  $\Delta=-0.20,-0.10,0.00,$ and $+0.10$, from top to bottom. Curves
  represent $\log_{10}(m_\phi/{\rm eV})=-18,-20,-22,-24,-25,-26,$
  and $-27$, as identified in the legend. Lower scalar-field masses
  produce stronger suppression that extends toward larger physical
  scales.}
  \label{fig:sfdm_wvcdm_pk}
\end{figure*}

\section{Results and discussion}
\label{sec:results_discussion}

We first investigate the theoretical dependence of the linear matter power spectrum and CMB temperature power spectrum on the scalar-field mass and the dark energy transition parameters.

An important characteristic of SFDM is its
scale-dependent suppression of matter fluctuations.
To quantify the deviations from the reference
$\Lambda$CDM cosmology, we consider the fractional
difference in the linear matter power spectrum,
defined as
\begin{equation}
R(k) = \frac{P_{\rm model}(k)}{P_{\Lambda\mathrm{CDM}}(k)}-1.
  \label{eq:pk_fractional_results}
\end{equation}

Here $P_{\rm model}(k)$ denotes the linear matter
power spectrum predicted by SFDM+$w_\dagger$VCDM,
while $P_{\Lambda\mathrm{CDM}}(k)$ represents the
corresponding reference spectrum. Figure~\ref{fig:sfdm_wvcdm_pk}
presents this fractional difference for representative
scalar-field masses
$\log_{10}(m_\phi/{\rm eV})=-18,-20,-22,-24,-25,-26,$
and $-27$. The two columns correspond to transition
redshifts $z_\dagger=0.30$ and $0.667$, while the
four rows show $\Delta=-0.20,-0.10,0.00,$ and $+0.10$.

For the heavier scalar-field masses, the predictions
remain close to the $\Lambda$CDM reference over the
displayed wavenumber range. As the scalar-field mass
decreases, the suppression becomes progressively more
pronounced and extends toward smaller wavenumbers,
corresponding to larger physical scales. In particular,
the models with $m_\phi=10^{-26}\,{\rm eV}$ and
$10^{-27}\,{\rm eV}$ exhibit substantial departures
from the reference spectrum, with the lightest mass
approaching nearly complete suppression at sufficiently
large displayed wavenumbers. This behavior reflects
the scale-dependent effective pressure associated
with scalar-field perturbations, which suppresses
gravitational clustering on scales smaller than
the characteristic Jeans length. Comparing the
different rows and columns further reveals that
the scalar-field mass primarily determines the
scale and amplitude of the suppression, whereas
variations in the late-time transition parameters
produce comparatively smaller modifications to
the displayed spectra.

\begin{figure*}[tbp]
  \centering
\includegraphics[width=0.91\textwidth,height=0.76\textheight,keepaspectratio]{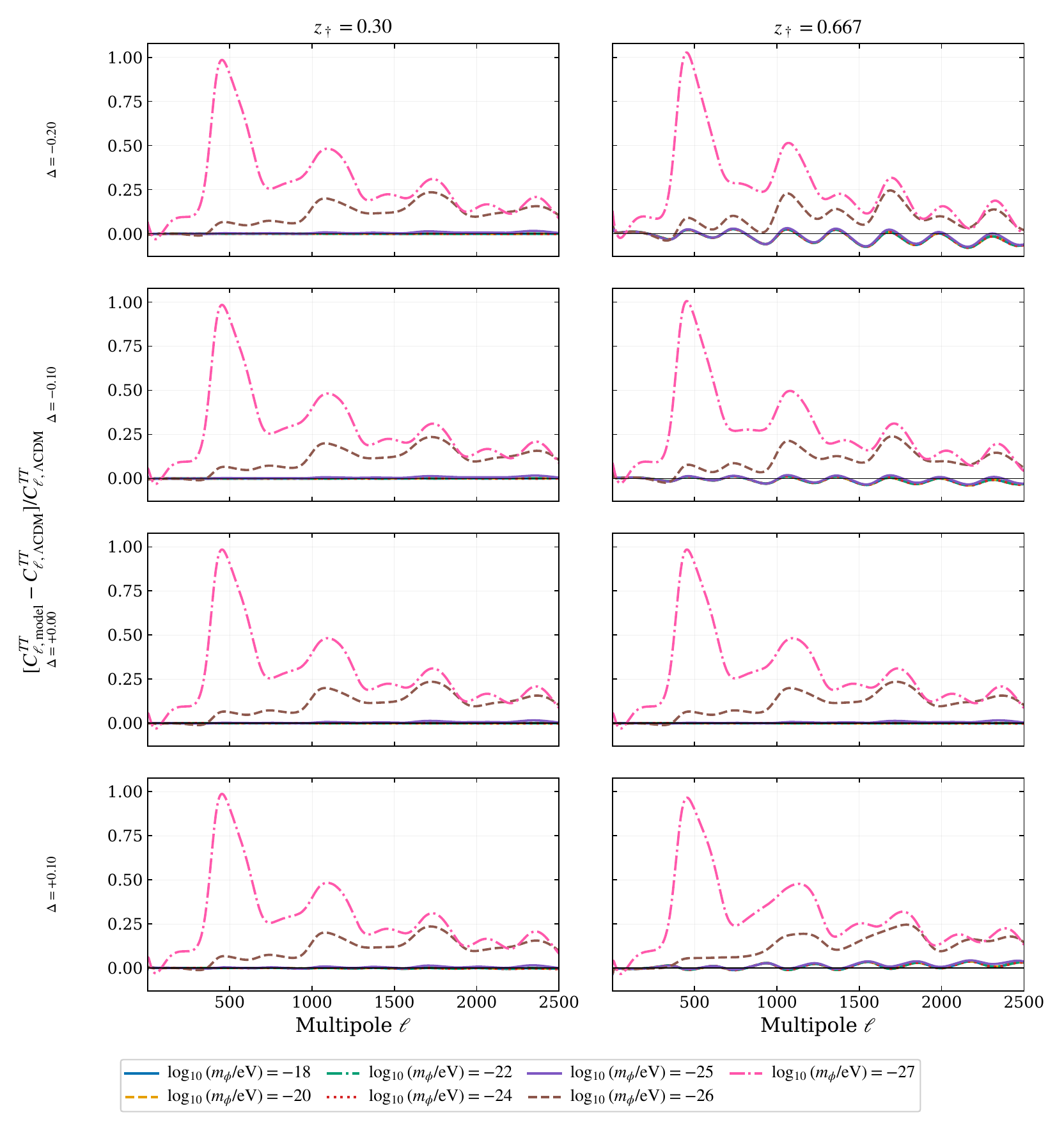}
 \caption{Fractional difference in the CMB temperature anisotropy
  spectrum, $[C_{\ell,\rm model}^{TT}-
  C_{\ell,\Lambda\mathrm{CDM}}^{TT}]/
  C_{\ell,\Lambda\mathrm{CDM}}^{TT}$, for SFDM+$w_\dagger$VCDM relative
  to $\Lambda$CDM. The columns show $z_\dagger=0.30$ and $0.667$,
  and the rows correspond to $\Delta=-0.20,-0.10,0.00,$ and $+0.10$.
  The line styles and colors identify scalar-field masses from
  $\log_{10}(m_\phi/{\rm eV})=-18$ to $-27$. Heavier masses yield
  smaller fractional differences, while the lightest cases produce
  substantial scale-dependent oscillatory features. The curves are
  illustrative theoretical predictions, not posterior confidence
  regions.}
  \label{fig:sfdm_wvcdm_tt}
\end{figure*}

We next examine the impact of the same benchmark
models on the CMB temperature anisotropies. To
characterize the corresponding deviations from
$\Lambda$CDM, we consider the fractional difference
in the temperature angular power spectrum,
\begin{equation}
R(\ell) = \frac{C_{\ell,\rm model}^{TT}}
       {C_{\ell,\Lambda\mathrm{CDM}}^{TT}}-1.
  \label{eq:tt_fractional_results}
\end{equation}

Figure~\ref{fig:sfdm_wvcdm_tt}
shows this quantity for the same scalar-field masses
and transition-parameter configurations considered
in Fig.~\ref{fig:sfdm_wvcdm_pk}.

For the heavier scalar-field masses, the fractional
differences remain small across most of the
displayed multipole range, indicating that their
CMB temperature spectra closely approach the
$\Lambda$CDM prediction. In contrast, decreasing
the scalar-field mass produces increasingly
pronounced deviations, including oscillatory
features and changes in the amplitudes of the
acoustic peaks. These effects are particularly
prominent for
$\log_{10}(m_\phi/{\rm eV})=-26$ and $-27$.
The substantial deviations associated with the
lightest benchmark demonstrate that sufficiently
light scalar fields can affect perturbations on
scales relevant to the CMB temperature anisotropies.
However, the magnitude of these theoretical
departures alone does not establish their
observational viability, which must be assessed
through the corresponding likelihood analysis.

The relative agreement of the heavier-mass models
with $\Lambda$CDM illustrates the reduced
sensitivity of these observables to the scalar-field
mass as the cold-dark-matter limit is approached.
Consequently, the CMB and background-distance
measurements may allow a broad range of sufficiently
large scalar-field masses. In comparison, the linear
matter power spectrum provides a direct illustration
of the mass-dependent suppression of density
perturbations as a function of comoving wavenumber.

\begin{table*}[t]
\caption{Marginalized constraints at 68\% confidence level for the free
and selected derived parameters of the models considered in this work,
based on CMB data combined with \textit{BAO DESI DR2}, with and
without Pantheon+.} 
\label{tab:comparison}
\begin{ruledtabular}
\begin{tabular}{llcc}
Parameter & Model & \textit{Planck+BAO DESI DR2+Pantheon+} & \textit{Planck+BAO DESI DR2} \\
\hline
$10^2\omega_b$
& $w_\dagger$CDM
& $2.249\pm0.012$ & $2.241\pm0.013$ \\
& {\color{blue}$w_\dagger$VCDM}
& {\color{blue}$2.250\pm0.012$}
& {\color{blue}$2.244\pm0.013$} \\
& {\color{red!70!black}SFDM+$w_\dagger$VCDM}
& {\color{red!70!black}$2.2491\pm0.0126$}
& {\color{red!70!black}$2.2415\pm0.0131$} \\[2pt]

$\omega_{\rm cdm}$
& $w_\dagger$CDM
& $0.11836\pm0.00058$ & $0.11929\pm0.00070$ \\
& {\color{blue}$w_\dagger$VCDM}
& {\color{blue}$0.11835\pm0.00054$}
& {\color{blue}$0.11922^{+0.00079}_{-0.00070}$} \\
& {\color{red!70!black}SFDM+$w_\dagger$VCDM}
& --- & --- \\[2pt]

$\Omega_{\phi,0}$
& $w_\dagger$CDM & --- & --- \\
& {\color{blue}$w_\dagger$VCDM} & --- & --- \\
& {\color{red!70!black}SFDM+$w_\dagger$VCDM}
& {\color{red!70!black}$0.2605\pm0.0051$}
& {\color{red!70!black}$0.2788\pm0.0132$} \\[2pt]

$100\theta_s$
& $w_\dagger$CDM
& $1.04206\pm0.00028$ & $1.04198\pm0.00025$ \\
& {\color{blue}$w_\dagger$VCDM}
& {\color{blue}$1.04205\pm0.00027$}
& {\color{blue}$1.04196\pm0.00027$} \\
& {\color{red!70!black}SFDM+$w_\dagger$VCDM}
& {\color{red!70!black}$1.04203\pm0.00028$}
& {\color{red!70!black}$1.04195\pm0.00047$} \\[2pt]

$\ln(10^{10}A_s)$
& $w_\dagger$CDM
& $3.052\pm0.015$ & $3.049\pm0.013$ \\
& {\color{blue}$w_\dagger$VCDM}
& {\color{blue}$3.050^{+0.013}_{-0.015}$}
& {\color{blue}$3.045\pm0.014$} \\
& {\color{red!70!black}SFDM+$w_\dagger$VCDM}
& {\color{red!70!black}$3.0466\pm0.0159$}
& {\color{red!70!black}$3.0444\pm0.0156$} \\[2pt]

$n_s$
& $w_\dagger$CDM
& $0.9692\pm0.0033$ & $0.9672\pm0.0032$ \\
& {\color{blue}$w_\dagger$VCDM}
& {\color{blue}$0.9693\pm0.0032$}
& {\color{blue}$0.9675\pm0.0034$} \\
& {\color{red!70!black}SFDM+$w_\dagger$VCDM}
& {\color{red!70!black}$0.9693\pm0.0035$}
& {\color{red!70!black}$0.9668\pm0.0037$} \\[2pt]

$\tau_{\rm reio}$
& $w_\dagger$CDM
& $0.0594^{+0.0067}_{-0.0077}$
& $0.0567^{+0.0059}_{-0.0068}$ \\
& {\color{blue}$w_\dagger$VCDM}
& {\color{blue}$0.0584^{+0.0064}_{-0.0077}$}
& {\color{blue}$0.0555^{+0.0075}_{-0.0067}$} \\
& {\color{red!70!black}SFDM+$w_\dagger$VCDM}
& {\color{red!70!black}$0.0575\pm0.0076$}
& {\color{red!70!black}$0.0553\pm0.0076$} \\[2pt]

$\Delta$
& $w_\dagger$CDM
& $-0.102^{+0.031}_{-0.035}$ & $-0.226\pm0.070$ \\
& {\color{blue}$w_\dagger$VCDM}
& {\color{blue}$-0.103\pm0.031$}
& {\color{blue}$-0.220\pm0.077$} \\
& {\color{red!70!black}SFDM+$w_\dagger$VCDM}
& {\color{red!70!black}$-0.0888\pm0.0330$}
& {\color{red!70!black}$-0.2304\pm0.0981$} \\[2pt]

$z_\dagger$
& $w_\dagger$CDM
& $0.553^{+0.047}_{-0.140}$ & $0.533\pm0.064$ \\
& {\color{blue}$w_\dagger$VCDM}
& {\color{blue}$0.504\pm0.095$}
& {\color{blue}$0.521^{+0.057}_{-0.052}$} \\
& {\color{red!70!black}SFDM+$w_\dagger$VCDM}
& {\color{red!70!black}$0.527\pm0.144$}
& {\color{red!70!black}$0.5127\pm0.0830$} \\
\hline

$H_0\,[{\rm km\,s^{-1}Mpc^{-1}}]$
& $w_\dagger$CDM
& $67.13\pm0.55$ & $65.3\pm1.1$ \\
& {\color{blue}$w_\dagger$VCDM}
& {\color{blue}$67.18\pm0.53$}
& {\color{blue}$65.4\pm1.1$} \\
& {\color{red!70!black}SFDM+$w_\dagger$VCDM}
& {\color{red!70!black}$67.32\pm0.57$}
& {\color{red!70!black}$65.37\pm1.39$} \\[2pt]

$\Omega_m$
& $w_\dagger$CDM
& $0.3141\pm0.0058$ & $0.335\pm0.012$ \\
& {\color{blue}$w_\dagger$VCDM}
& {\color{blue}$0.3136\pm0.0054$}
& {\color{blue}$0.333\pm0.013$} \\
& {\color{red!70!black}SFDM+$w_\dagger$VCDM}
& {\color{red!70!black}$0.3116\pm0.0059$}
& {\color{red!70!black}$0.3327\pm0.0153$} \\[2pt]

$S_8$
& $w_\dagger$CDM
& $0.8077\pm0.0083$ & $0.824\pm0.011$ \\
& {\color{blue}$w_\dagger$VCDM}
& {\color{blue}$0.8068\pm0.0080$}
& {\color{blue}$0.821^{+0.012}_{-0.011}$} \\
& {\color{red!70!black}SFDM+$w_\dagger$VCDM}
& {\color{red!70!black}$0.8148\pm0.0110$}
& {\color{red!70!black}$0.8319\pm0.0156$} \\
\hline

$\log_{10}(m_\phi/{\rm eV})$ (95\% lower limit)
& $w_\dagger$CDM & --- & --- \\
& {\color{blue}$w_\dagger$VCDM} & --- & --- \\
& {\color{red!70!black}SFDM+$w_\dagger$VCDM}
& {\color{red!70!black}$>-24.04$}
& {\color{red!70!black}$>-24.11$} \\
\hline
\end{tabular}
\end{ruledtabular}
\end{table*}

\subsection{Observational Constrains}

\begin{figure*}[tbp]
  \centering
\includegraphics[width=0.92\textwidth,height=0.78\textheight,keepaspectratio]{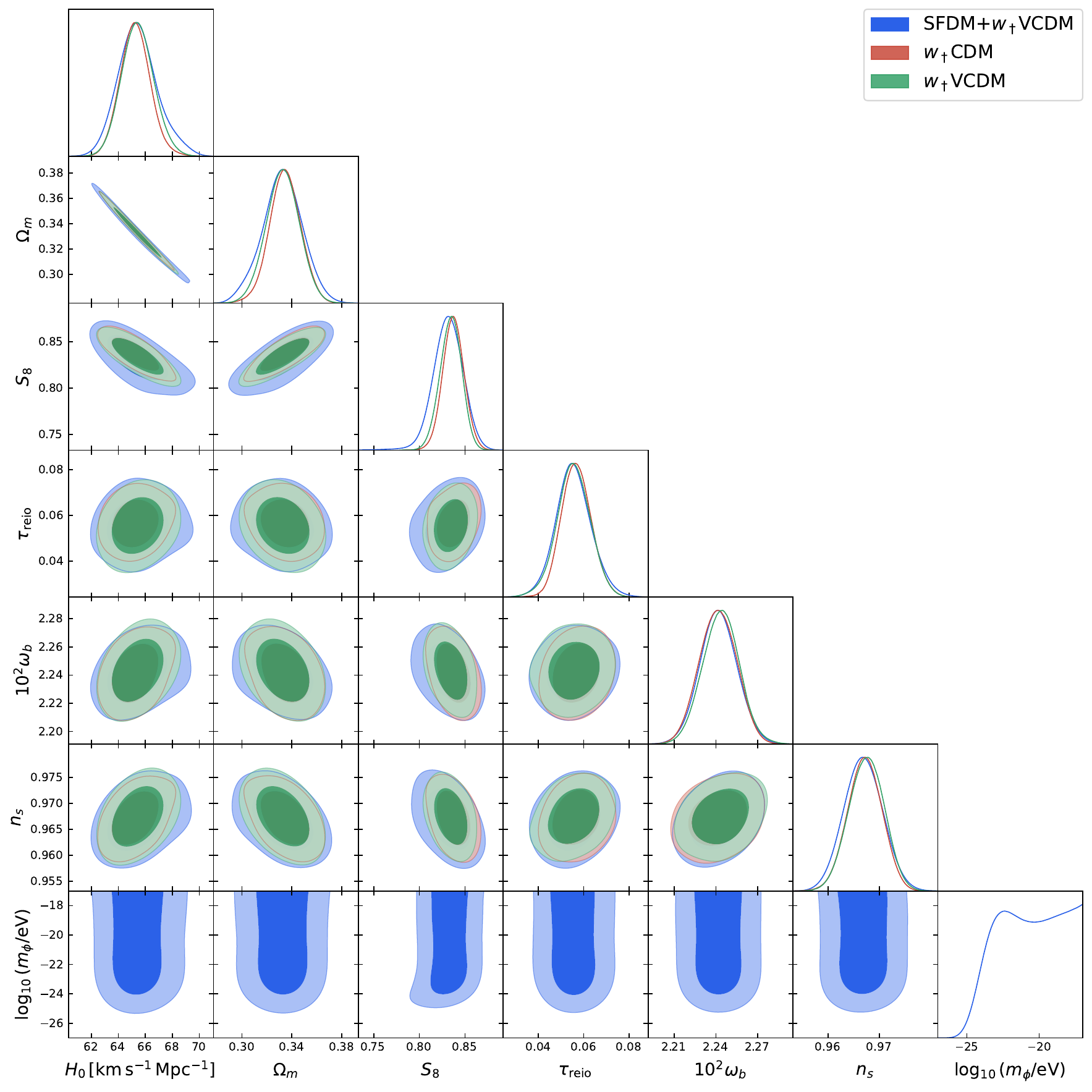}
        \caption{Marginalized one-dimensional posterior distributions and
  two-dimensional credible contours for $H_0$, $\Omega_m$, $S_8$,
  $\tau_{\rm reio}$, $10^2\omega_b$, $n_s$, and
  $\log_{10}(m_\phi/{\rm eV})$, using \textit{Planck} and \textit{BAO DESI DR2}
  data. Blue, red, and green denote SFDM+$w_\dagger$VCDM,
  $w_\dagger$CDM, and $w_\dagger$VCDM, respectively. The scalar-field
  mass appears only for SFDM+$w_\dagger$VCDM. Inner and outer shaded
  contours show the plotted 68\% and 95\% marginalized credible
  regions.}
  \label{fig:sfdm_wvcdm_seven_dr2}
\end{figure*}

\begin{figure*}[tbp]
  \centering
\includegraphics[width=0.92\textwidth,height=0.78\textheight,keepaspectratio]{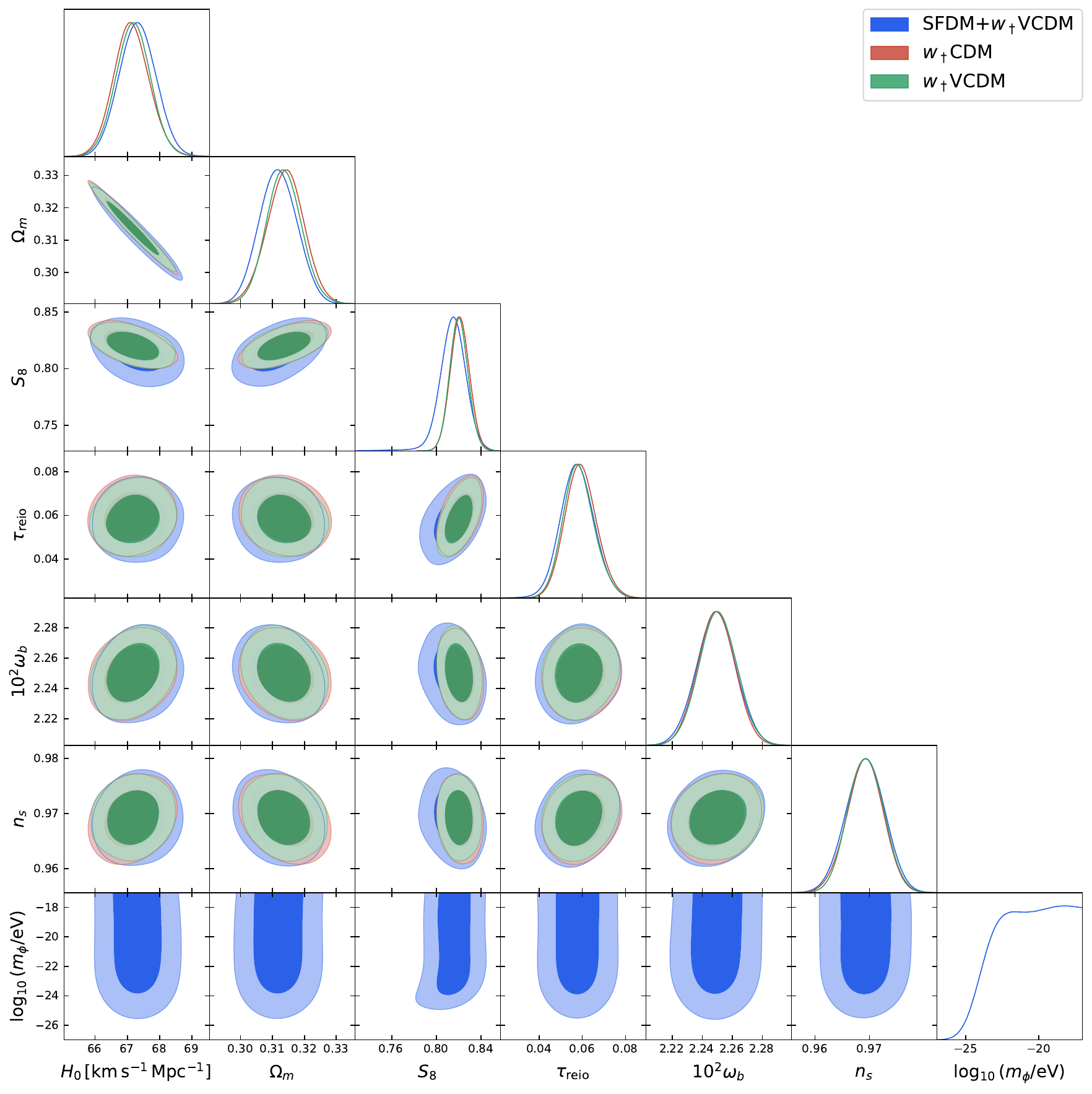}
        \caption{Same seven-parameter posterior comparison as in
  Fig.~\ref{fig:sfdm_wvcdm_seven_dr2}, now including Pantheon+
  supernovae in addition to \textit{Planck}  and \textit{BAO DESI DR2}. The three
  model colors are unchanged. The figure illustrates how adding
  supernova distances changes the inferred expansion and clustering
  parameters, while the SFDM scalar-field mass retains a broad
  high-mass posterior.}
  \label{fig:sfdm_wvcdm_seven_dr2pp}
\end{figure*}

We now turn to the MCMC parameter constraints. Table~\ref{tab:comparison} summarizes the marginalized posterior estimates for the three cosmological models under the two observational data combinations, while Figs.~\ref{fig:sfdm_wvcdm_seven_dr2} and \ref{fig:sfdm_wvcdm_seven_dr2pp} show the corresponding joint posterior distributions. The SFDM+$w_\dagger$VCDM model is represented in blue, whereas $w_\dagger$CDM and $w_\dagger$VCDM are shown in red and green, respectively. The parameter space includes $H_0$, $\Omega_m$, $S_8$, $\tau_{\rm reio}$, $10^2\omega_b$, and $n_s$, with $\log_{10}(m_\phi/{\rm eV})$ as an additional parameter specific to the SFDM scenario.

For the \textit{Planck+BAO DESI DR2} data combination, the three models exhibit substantial overlap in the posterior distributions of the standard cosmological parameters, indicating broadly consistent constraints across the different theoretical frameworks. Nevertheless, SFDM+$w_\dagger$VCDM yields broader posterior distributions along several parameter directions, while preserving the characteristic parameter degeneracies established in previous studies. In particular, a pronounced anticorrelation between $H_0$ and $\Omega_m$ is observed, reflecting the well-known degeneracy between the present-day expansion rate and the matter density. The joint posterior distributions involving $S_8$ further illustrate the interplay between the matter density and the amplitude of matter clustering, as well as the dependence of the inferred clustering constraints on the underlying cosmological model.

The inclusion of Pantheon+ provides complementary luminosity-distance
information and shifts and tightens several parameter distributions
relative to \textit{Planck+BAO DESI DR2}, as shown in
Fig.~\ref{fig:sfdm_wvcdm_seven_dr2pp}. For SFDM+$w_\dagger$VCDM,
we find 
$H_0=65.37\pm1.39\,{\rm km\,s^{-1}\,Mpc^{-1}}$ and
$\Omega_m=0.3327\pm0.0153$ for \textit{Planck+BAO DESI DR2}, compared with
$H_0=67.32\pm0.57\,{\rm km\,s^{-1}\,Mpc^{-1}}$ and
$\Omega_m=0.3116\pm0.0059$ after adding Pantheon+. Thus the
Pantheon+ combination favors a higher $H_0$ and lower $\Omega_m$ in the
SFDM scenario, with smaller quoted marginalized uncertainties as expected from \cite{PantheonPlus}'s results. The
corresponding constraints on $S_8\equiv\sigma_8\sqrt{\Omega_m/0.3}$
change from $0.8319\pm0.0156$ to $0.8148\pm0.0110$. The standard cosmological parameters remain mutually compatible across
the three fitted models at the level illustrated by their overlapping
posterior contours. 

In the SFDM model the present-day scalar-field density fraction changes from
$\Omega_{\phi,0}=0.2788\pm0.0132$ to
$0.2605\pm0.0051$. The absence of an $\omega_{\rm cdm}$ entry for this
model in Table~\ref{tab:comparison} reflects the replacement of the CDM fluid by the scalar field, rather than a missing measurement.

\begin{figure*}[tbp]
  \centering
  \includegraphics[width=0.48\textwidth,height=0.48\textheight,keepaspectratio]{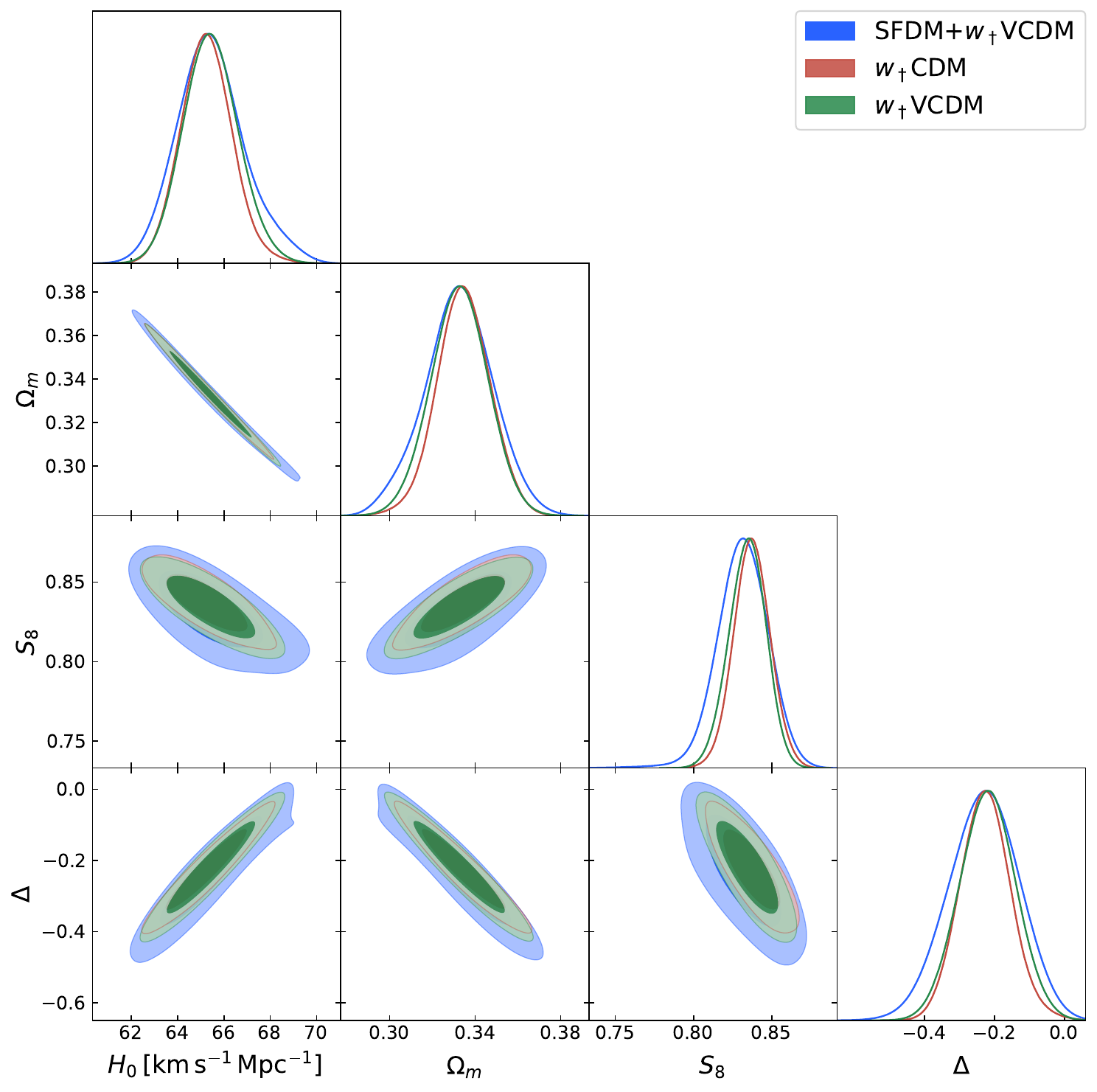} \,\,\,\,
  \includegraphics[width=0.48\textwidth,height=0.48\textheight,keepaspectratio]{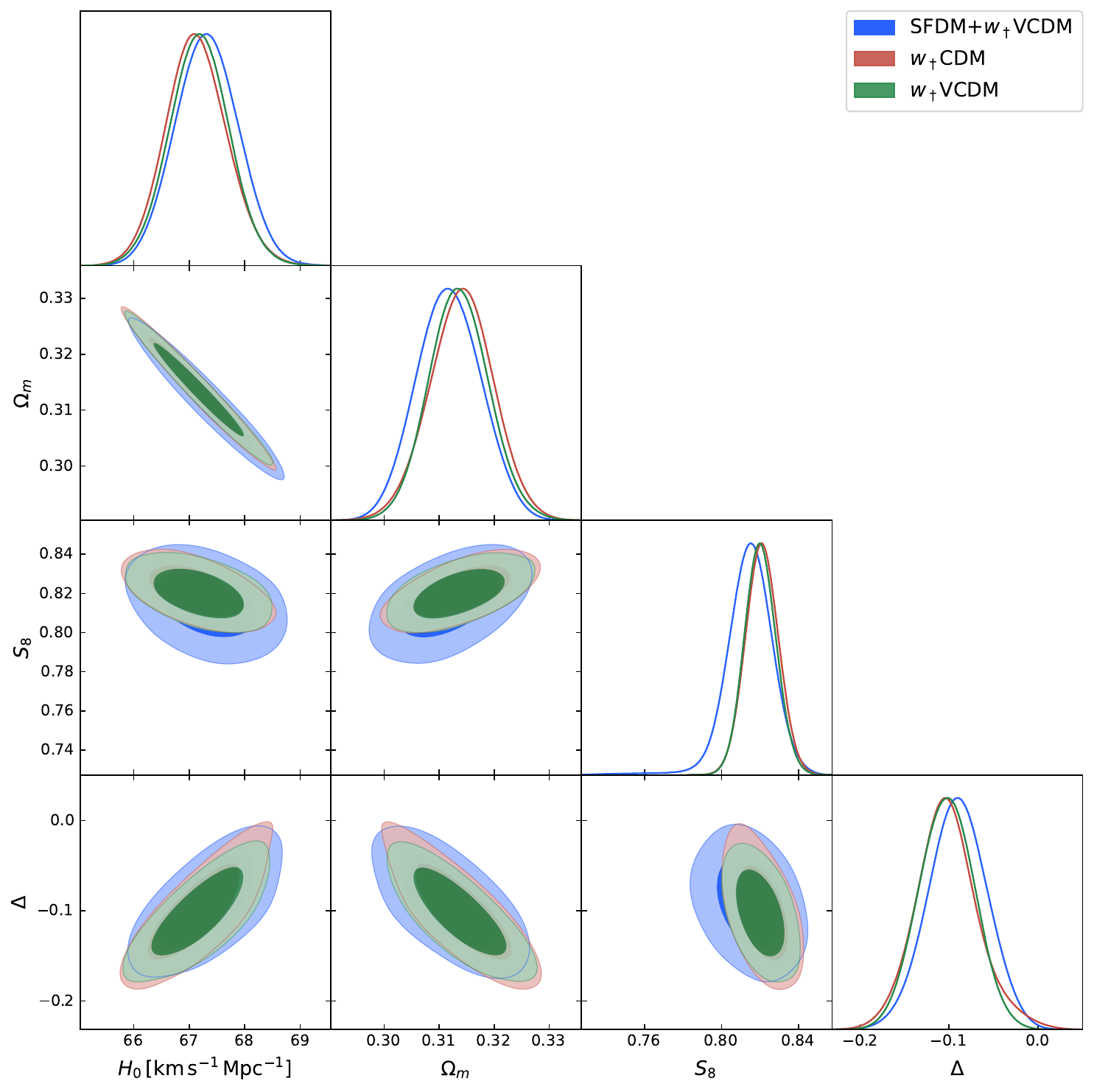}
        \caption{Left panel: Marginalized posterior distributions and joint constraints on the four parameters $H_0$, $\Omega_m$, $S_8$, and $\Delta$, obtained from \textit{Planck} CMB and DESI DR2 BAO data. The blue, red, and green contours correspond to SFDM+$w_\dagger$VCDM, $w_\dagger$CDM, and $w_\dagger$VCDM, respectively. The 1D curves show the marginalized posterior distributions, while the inner and outer contours indicate the 68\% and 95\% credible regions, respectively. Correlations involving $\Delta$ illustrate the interplay between the late-time transition and the inferred cosmological parameters. Right panel: Same as the left panel, but including the Pantheon+ Type Ia supernova data in the joint analysis of \textit{Planck} CMB, DESI DR2 BAO, and Pantheon+ observations.}
  \label{fig:sfdm_wvcdm_four_dr2}
\end{figure*}

The comparatively simple four-parameter projections in
Figs.~\ref{fig:sfdm_wvcdm_four_dr2} isolate $H_0$, $\Omega_m$, $S_8$, and
the transition-amplitude parameter $\Delta$. In both data combinations,
$H_0$ and $\Omega_m$ exhibit an elongated anticorrelation. The contours
involving $\Delta$ demonstrate its covariance with the inferred
expansion parameters: different combinations of the transition amplitude
and background parameters can reproduce similar integrated distance
constraints. For \textit{Planck+BAO DESI DR2}, the SFDM+$w_\dagger$VCDM marginalized
constraint is $\Delta=-0.2304\pm0.0981$, while the two comparison
models give $\Delta=-0.226\pm0.070$ ($w_\dagger$CDM) and
$-0.220\pm0.077$ ($w_\dagger$VCDM). 

\begin{figure*}[tbp]
  \centering
  \includegraphics[width=0.48\textwidth,height=0.48\textheight,keepaspectratio]{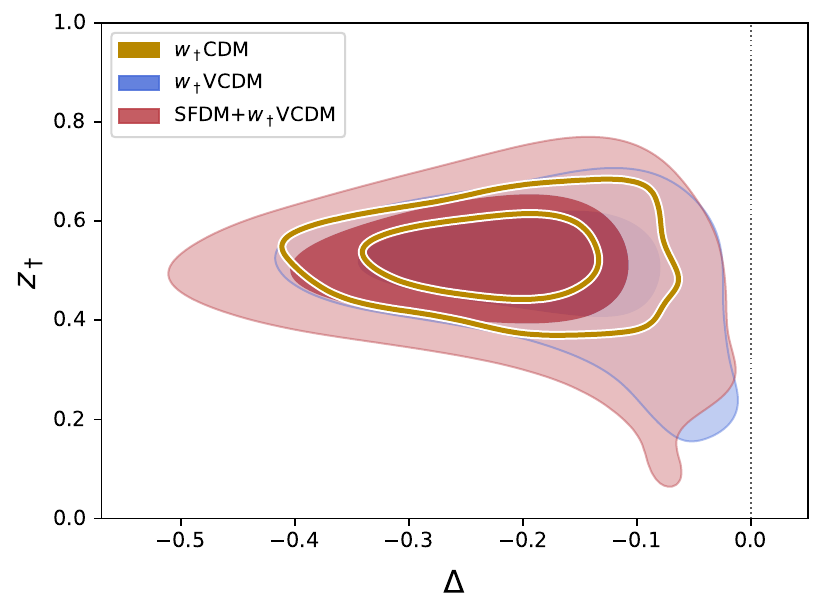} 
   \includegraphics[width=0.48\textwidth,height=0.48\textheight,keepaspectratio]{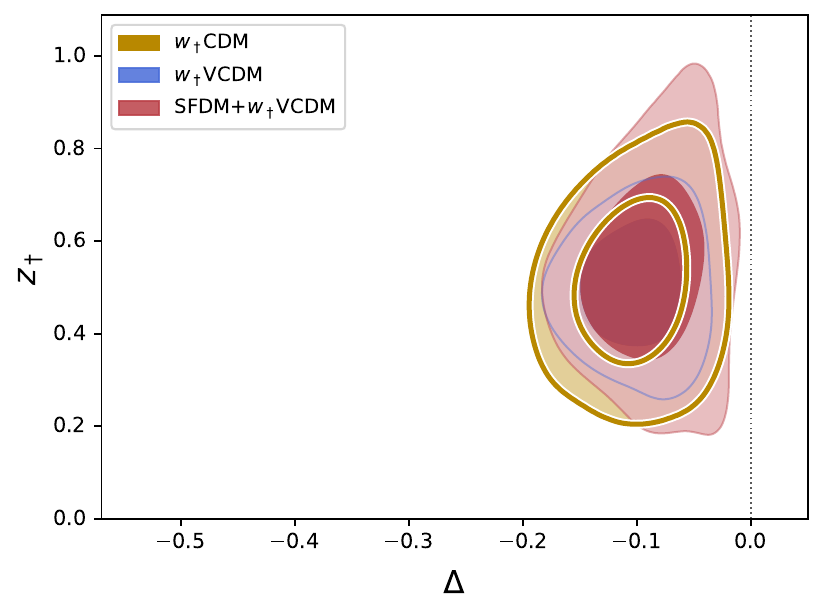}
  \caption{Left panel: Parameter constraints in the $(\Delta,z_\dagger)$ plane for $w_\dagger$CDM, $w_\dagger$VCDM, and SFDM+$w_\dagger$VCDM, obtained from the combination of \textit{Planck} and \textit{BAO DESI DR2} data. The shaded regions indicate the 68\% and 95\% Bayesian credible regions. The vertical dotted line marks the $\Delta=0$ limit. The orientation and extent of the contours illustrate the degeneracy between the transition amplitude, $\Delta$, and the transition redshift, $z_\dagger$. Right panel: Same as the left panel, but including the Pantheon+ Type Ia supernova data in the joint analysis of \textit{Planck} CMB, DESI DR2 BAO, and Pantheon+ observations.}
  \label{fig:sfdm_wvcdm_delta_z_dr2}
\end{figure*}

Including Pantheon+ data yields a transition-amplitude constraint of $\Delta=-0.0888\pm0.0330$ for SFDM+$w_\dagger$VCDM, compared with $\Delta=-0.102^{+0.031}_{-0.035}$ for $w_\dagger$CDM and $\Delta=-0.103\pm0.031$ for $w_\dagger$VCDM. As shown in Fig.~\ref{fig:sfdm_wvcdm_delta_z_dr2}, the inclusion of Pantheon+ shifts the posterior distribution of $\Delta$ toward zero and leads to a narrower constraint relative to the \textit{Planck+BAO DESI DR2}-only analysis. The posterior distributions of $H_0$ and $\Omega_m$ also exhibit shifts consistent with those quantified above. These results demonstrate that late-time distance measurements from Type Ia supernovae provide additional constraining power on the transition amplitude, modifying the inferred parameter constraints rather than merely reducing the uncertainties associated with the CMB+BAO analysis.


The transition redshift, $z_\dagger$, is constrained jointly with the transition amplitude, $\Delta$, as shown in Figs.~\ref{fig:sfdm_wvcdm_delta_z_dr2}. For SFDM+$w_\dagger$VCDM, the \textit{Planck+BAO DESI DR2} combination yields $z_\dagger=0.5127\pm0.0830$ and $\Delta=-0.2304\pm0.0981$. Upon including Pantheon+, the constraints become $z_\dagger=0.527\pm0.144$ and $\Delta=-0.0888\pm0.0330$. Thus, the central value of the transition redshift remains close to $z_\dagger\simeq0.5$, while the inferred magnitude of the transition amplitude decreases substantially when Pantheon+ data are included. A negative $\Delta$ corresponds, within the adopted parameterization, to phantom-like effective dark-energy behavior ($w<-1$) before the transition and quintessence-like behavior ($w>-1$) afterward. The resulting constraints therefore favor this sequence of effective DE regimes.

The additional SFDM mass parameter distinguishes the scalar-field model from both comparison frameworks and determines the characteristic scale below which pressure effects associated with field gradients suppress the growth of cosmological perturbations. As shown in the last row and diagonal panel of Figs.~\ref{fig:sfdm_wvcdm_seven_dr2} and \ref{fig:sfdm_wvcdm_seven_dr2pp}, the posterior distribution of $m_\phi$ is broad and extends toward the high-mass end of the explored prior, without exhibiting a sharply localized maximum that would identify a preferred mass scale. This behavior is consistent with the expected phenomenology of ultralight scalar-field dark matter: increasing the field mass shifts the suppression of structure formation toward smaller scales, progressively reducing departures from the standard CDM predictions on cosmological scales.

As illustrated by Figs.~\ref{fig:sfdm_wvcdm_pk} and \ref{fig:sfdm_wvcdm_tt}, sufficiently small masses lead to substantial modifications in the linear matter power spectrum and the CMB temperature anisotropy spectrum, respectively. These deviations provide a physical explanation for the reduced likelihood in the low-mass regime. Conversely, at higher masses, the scalar-field dynamics increasingly approach the CDM limit over the range of scales probed by the data, reducing the ability of these observations to distinguish between the two descriptions. The resulting posterior can therefore exhibit a lower-mass exclusion without a well-defined upper constraint or a localized mass determination. Degeneracies with other cosmological parameters governing the background expansion and perturbation evolution may further broaden the allowed mass range.

Using the one-sided 95\% credible limits reported in Table~\ref{tab:comparison}, we obtain
\begin{align}
  \log_{10}(m_\phi/{\rm eV}) &> -24.11
  &&\textit{(Planck + BAO DESI DR2)},\nonumber\\
  \log_{10}(m_\phi/{\rm eV}) &> -24.04
  &&\textit{(Planck + BAO DESI DR2 + PP)}.
  \label{eq:sfdm_mass_limits_results}
\end{align}

\begin{figure*}[tbp]
  \centering
  \includegraphics[width=0.97\textwidth,height=0.67\textheight,keepaspectratio]{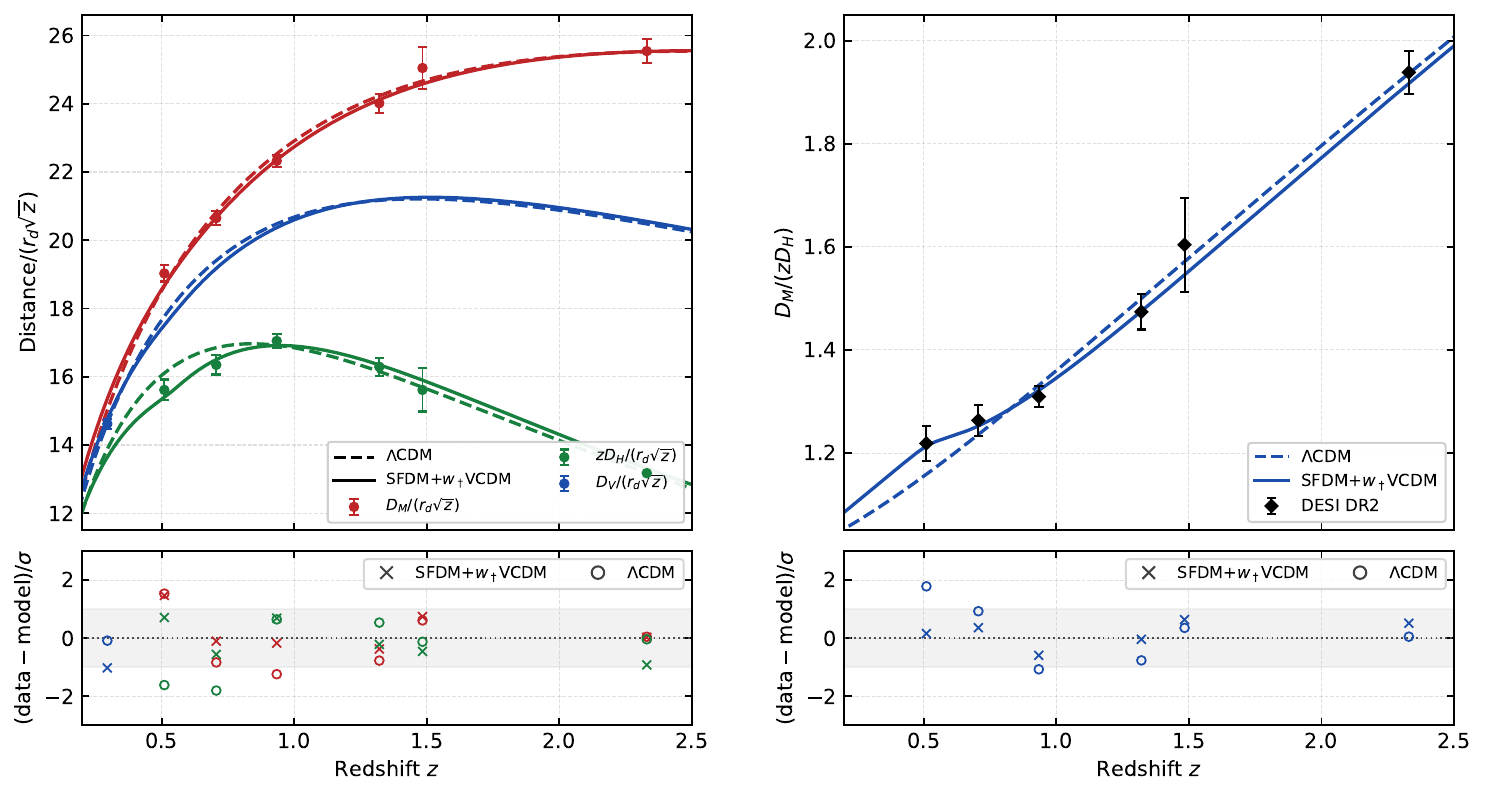}
    \caption{Comparison of SFDM+$w_\dagger$VCDM and $\Lambda$CDM predictions with \textit{BAO DESI DR2} measurements for the \textit{Planck+BAO DESI DR2} analysis. The upper panels show the rescaled distances $D_M/(r_d\sqrt{z})$, $zD_H/(r_d\sqrt{z})$, and $D_V/(r_d\sqrt{z})$ (left), and the Alcock--Paczynski ratio $D_M/(zD_H)$ (right), where $D_H=c/H(z)$. Solid and dashed curves represent SFDM+$w_\dagger$VCDM and $\Lambda$CDM, respectively. The lower panels show normalized residuals, with crosses and circles denoting the two models, respectively. The gray band indicates the approximate $\pm1\sigma$ region.}
  \label{fig:sfdm_wvcdm_bao_dr2}
\end{figure*}

The inclusion of Pantheon+ shifts the lower bound slightly toward higher masses. Given the modest difference between these limits, the addition of supernova data does not substantially improve the mass constraint. Instead, the results indicate that the inferred lower bound remains broadly stable under the inclusion of additional geometrical information.

These constraints are conditional on the assumed SFDM+$w_\dagger$VCDM framework, the adopted mass prior, and the cosmological likelihoods considered here. In particular, they should not be directly identified with bounds derived from observables that probe the suppression of structure formation on smaller scales, such as the Lyman-$\alpha$ forest, since the resulting constraints depend on the observables, model assumptions, and statistical methodology. The present analysis therefore establishes a lower bound on the scalar-field mass within the adopted cosmological framework, rather than identifying a preferred ultralight mass scale or providing a model-independent determination of $m_\phi$.

We finally examine the background-distance predictions in Fig.~\ref{fig:sfdm_wvcdm_bao_dr2}, comparing SFDM+$w_\dagger$VCDM and $\Lambda$CDM with the \textit{BAO DESI DR2} measurements. The left panel shows the rescaled distances $D_M/(r_d\sqrt{z})$, $zD_H/(r_d\sqrt{z})$, and $D_V/(r_d\sqrt{z})$, where $D_H=c/H(z)$, while the right panel displays the dimensionless Alcock--Paczynski ratio $D_M/(zD_H)$. The lower panels show the normalized residuals in units of the observational uncertainties.

At $z>1$, both models exhibit similar trends and provide broadly comparable descriptions of the BAO measurements. At lower redshifts, their predictions diverge more noticeably around $z\sim0.5$, particularly for the radial distance, where SFDM+$w_\dagger$VCDM follows the observed BAO measurements more closely than $\Lambda$CDM. Smaller differences also appear in the transverse distance and the Alcock--Paczynski ratio. This low-redshift behavior is consistent with that reported for the $w_\dagger$CDM transition model in Ref.~\cite{Scherer2025}, indicating that incorporating quadratic scalar-field dark matter preserves the characteristic background signature of the late-time transition while introducing additional mass-dependent effects in the evolution of cosmological perturbations.

\section{Conclusions}
\label{sec:conclusions}

In this work, we investigated quadratic SFDM in a cosmological scenario featuring a smooth late-time dark-energy transition, motivated by recent observations of the CMB, BAO, and Type Ia supernovae. As expected, our linear-theory results show that the scalar-field mass determines the characteristic scale of departures from CDM clustering. Lower masses produce stronger suppression of the linear matter power spectrum and more pronounced modifications to the CMB temperature anisotropies, whereas higher masses recover nearly CDM-like predictions on the scales considered.

Using Planck CMB temperature and polarization data combined with BAO measurements, we constrained SFDM+$w_\dagger$VCDM and compared its predictions with those of $w_\dagger$CDM and $w_\dagger$VCDM. For the Planck+BAO DESI DR2 combination, we obtain $H_0=65.37\pm1.39\,\mathrm{km\,s^{-1}\,Mpc^{-1}}$, $\Omega_m=0.3327\pm0.0153$, and $S_8=0.8319\pm0.0156$ at 68\% credibility. The substantial overlap with the comparison models indicates that introducing quadratic SFDM does not significantly alter the preferred background cosmological parameters.

The transition parameters remain consistent with a late-time change in the effective dark-energy behavior, with $\Delta=-0.2304\pm0.0981$ and $z_\dagger=0.5127\pm0.0830$. Within the adopted parameterization, the negative transition amplitude corresponds to phantom-like behavior at higher redshifts and quintessence-like behavior at later times. 
For the scalar-field mass, we obtain the one-sided 95\% lower bound
\begin{equation}
  \log_{10}\!\left(\frac{m_\phi}{\mathrm{eV}}\right)>-24.11
\end{equation}
for \textit{Planck+BAO DESI DR2}. The posterior does not identify a preferred mass scale, reflecting the recovery of CDM-like predictions at higher masses. This result constrains the low-mass regime within the adopted model and prior, motivating complementary probes of small-scale structure.

Future observations will be essential to distinguish SFDM signatures from changes in the background expansion. Full-shape galaxy clustering, redshift-space distortions, weak lensing, and Lyman-$\alpha$ forest measurements can probe the scale-dependent suppression of structure and help break degeneracies among cosmological parameters. Complementarily, 21\,cm intensity mapping with the Square Kilometre Array Observatory (SKAO) will extend measurements of BAO and structure growth to broader redshift ranges \cite{weltman2020fundamental}. Upcoming and ongoing surveys will provide complementary constraints: the Nancy Grace Roman Space Telescope will combine galaxy clustering and weak lensing \cite{bartusek2022nancy}; the Legacy Survey of Space and Time (LSST) at the Rubin Observatory will deliver deep, wide-area imaging sensitive to structure growth and dark-matter physics \cite{LSSTScience:2009jmu}; Euclid will constrain dark matter and dark energy through galaxy clustering and weak lensing \cite{Euclid:2024yrr}; and the China Space Station Telescope (CSST) will combine weak lensing, galaxy clustering, and BAO \cite{csst}. Together, these observations will help determine whether the scale-dependent signatures of ultralight scalar-field dark matter can be distinguished from those of CDM-based cosmologies with a late-time transition.

\section{Data Availability}
\label{sec:data_availability}
Upon reasonable request, the corresponding author will disclose
the datasets utilized in this study.

\begin{acknowledgments}
MA and HIU gratefully acknowledge support from the SECIHTI PhD
scholarship program. This work received partial funding from
SECIHTI M\'exico under grants CBF-2025-G-1720 and CBF-2025-G-176.
Additional support was provided by grant I0101/131/07 C-234/07
of the Instituto Avanzado de Cosmolog\'ia (IAC) collaboration
(\url{http://www.iac.edu.mx/}). 
R.C.N. acknowledges financial support from the Conselho Nacional de Desenvolvimento Científico e Tecnológico (CNPq, National Council for Scientific and Technological Development) through Grant No. 304306/2022-3, and partial financial support from the Fundação de Amparo à Pesquisa do Estado do Rio Grande do Sul (FAPERGS, Research Support Foundation of the State of Rio Grande do Sul) through Grant No. 23/2551-0000848-3 and Grant No. 25/2551-0002612-1. 
PSF acknowledges support from the National Key Research and Development Program of China and the Zhejiang Provincial Top-Level Research Support Program. 
The authors acknowledge the computational resources provided by
LANCAD and SECIHTI through the Supercomputer Hybrid Cluster
``Xiuhcoatl'' at the General Coordination of Information and
Communication Technologies (CGSTIC) of CINVESTAV
(\url{http://clusterhibrido.cinvestav.mx/}).
MA thanks his fianc\'ee, Zunaira Shafique, for her emotional
support and encouragement throughout this study.
\end{acknowledgments}

\bibliographystyle{apsrev4-2}
\bibliography{biblio}

\end{document}